\documentclass{elsarticle}

\usepackage{url}
\usepackage{graphicx}
\usepackage{amsmath}
\usepackage{acronym}
\usepackage{multibib}
\usepackage{subcaption}
\usepackage{longtable}
\usepackage{array}
\usepackage{mdframed}
\usepackage{arydshln}
\usepackage{multirow}

\newcites{S}{Primary Studies}

\begin{document}
\newcommand{\etal}{ et al.}

\begin{frontmatter}

\title{A systematic mapping study of developer social network research}

\author{Steffen Herbold\corref{mycorrespondingauthor}\fnref{ifi}}
\cortext[mycorrespondingauthor]{Corresponding author}
\ead{herbold@cs.uni-goettingen.de}

\author{Aynur Amirfallah\fnref{ifi}}
\ead{aynur.amirfallah@stud.uni-goettingen.de}

\author{Fabian Trautsch\fnref{ifi}}
\ead{trautsch@cs.uni-goettingen.de}

\author{Jens Grabowski\fnref{ifi}}
\ead{grabowski@cs.uni-goettingen.de}

\address[ifi]{Institute of Computer Science, University of Goettingen, Germany}

\begin{abstract}
Developer social networks (DSNs) are a tool for the analysis of community structures and collaborations between developers in software projects and software ecosystems. Within this paper, we present the results of a systematic mapping study on the use of DSNs in software engineering research. We identified 255 primary studies on DSNs. We mapped the primary studies to research directions, collected information about the data sources and the size of the studies, and conducted a bibliometric assessment. We found that nearly half of the research investigates the structure of developer communities. Other frequent topics are prediction systems build using DSNs, collaboration behavior between developers, and the roles of developers. Moreover, we determined that many publications use a small sample size regarding the number of projects, which could be problematic for the external validity of the research. Our study uncovered several open issues in the state of the art, e.g., studying inter-company collaborations, using multiple information sources for DSN research, as well as general lack of reporting guidelines or replication studies. 
\end{abstract}


\end{frontmatter}

\acrodef{DSN}{Developer Social Network}
\acrodef{ITS}{Issue Tracking System}
\acrodef{VCS}{Version Control System}
\acrodef{ML}{Mailing List}

\section{Introduction}\label{sec:introduction}
Social structures within software development projects are a topic that received a lot of attention in different research communities, e.g., by researchers interested in open source development, global software engineering, and mining software repositories. \acp{DSN} are often inferred automatically from information that can be found in forges like GitHub, \acp{ML}, \acp{ITS}, and \acp{VCS} of software development projects. The \acp{DSN} give valuable insights into the projects, e.g., regarding the importance of individuals~\citeS{joblin2015developer}, patterns in communication behavior~\citeS{damian2007collaboration}, for the identification of single points of failure~\citeS{tamburri2019discovering}, gender-aspects~\citeS{catolino2019gender}, and even bugs~\citeS{pinzger2008can}. Due to the magnitude of publications on DSNs, the diversity of topics addressed by DSNs, and the lack of a contemporary literature review, a novel literature study is required to ensure that researchers and practitioners can get a complete overview on the state of the art of DSNs. This article describes a mapping study performed based on the rigorous guidelines by Kitchenham and Charters~\cite{Kitchenham2007} for literature reviews with the goal to identify and map research on \acp{DSN}. We map the publications on \acp{DSN} to research topics and analyze the scope of the publications in terms of data sources, number of projects, and number of people. 

With our mapping study, we provide the following contributions. 
\begin{itemize}
    \item A contemporary overview of the state of the art of the literature on \acp{DSN}. 
    \item A summary of the already investigated research directions, including the relevant literature. 
    \item A summary of the data sources, as well as the size of the \acp{DSN} in terms of number of projects and people involved. 
    \item A bibliometric assessment to identify influential publications, authors, venues, and interest in the topic over time.
    \item The identification of open issues within the current state of the art. 
\end{itemize}

We found that 49\% of all publications on \acp{DSN} analyze the structure of the community, either in general, or with respect to other aspects of software development, e.g., the evolution, or the impact on code quality. Other frequent topics in research are prediction systems based on \acp{DSN}, e.g., for defect prediction or bug triage, the collaboration behavior between developers, and the roles of developers. Regarding the way that studies are conducted, we found that 79\% of the studies are based on a single data source and 70\% of the studies use less then 11 projects to draw conclusions. These are concerning findings regarding the generalizability of results. Regardless, 80\% of publications use social networks with at least 100 people modelled by the network, i.e., large networks are usually the foundation for analysis, which is good for the generalizability. Thus, we believe there is a need for studies with high external validity on \acp{DSN}, especially more studies that consider a large amount of different projects in order to derive generalizable conclusions for diverse populations. Other open issues in the state of the art are, e.g.,  inter-company collaborations and the use of data from multiple information sources for the analysis of \acp{DSN}. Finally, the extraction of data from the publications for this mapping study revealed a lack of reporting guidelines for \acp{DSN}, i.e., some publications fail to report basic meta data about the studies conducted, e.g., the number of projects considered, the number of developers involved, or how data was processed, e.g., to deal with duplicate identities. 

The remainder of this paper is organized as follows. We give a definition of \acp{DSN} in Section~\ref{sec:foundations}. In Section~\ref{sec:methodology}, we present our methodology for the mapping study, including our research questions, inclusion and exclusion criteria for the literature, how we identified publications, and the data we collected for each included publication.  In Section~\ref{sec:review}, we give the results of our review, by listing the primary studies we found and map them to \ac{DSN} concepts according to our research questions. In Section~\ref{sec:discussion}, we discuss open issues regarding \ac{DSN} research based on the results of our mapping study. Then, we discuss related prior literature studies in Section~\ref{sec:relatedwork}, and conclude the article in Section~\ref{sec:conclusion}.

\section{Definition of \acfp{DSN}}
\label{sec:foundations}
\begin{figure}[htp]
\centering
\includegraphics[width=0.48\textwidth]{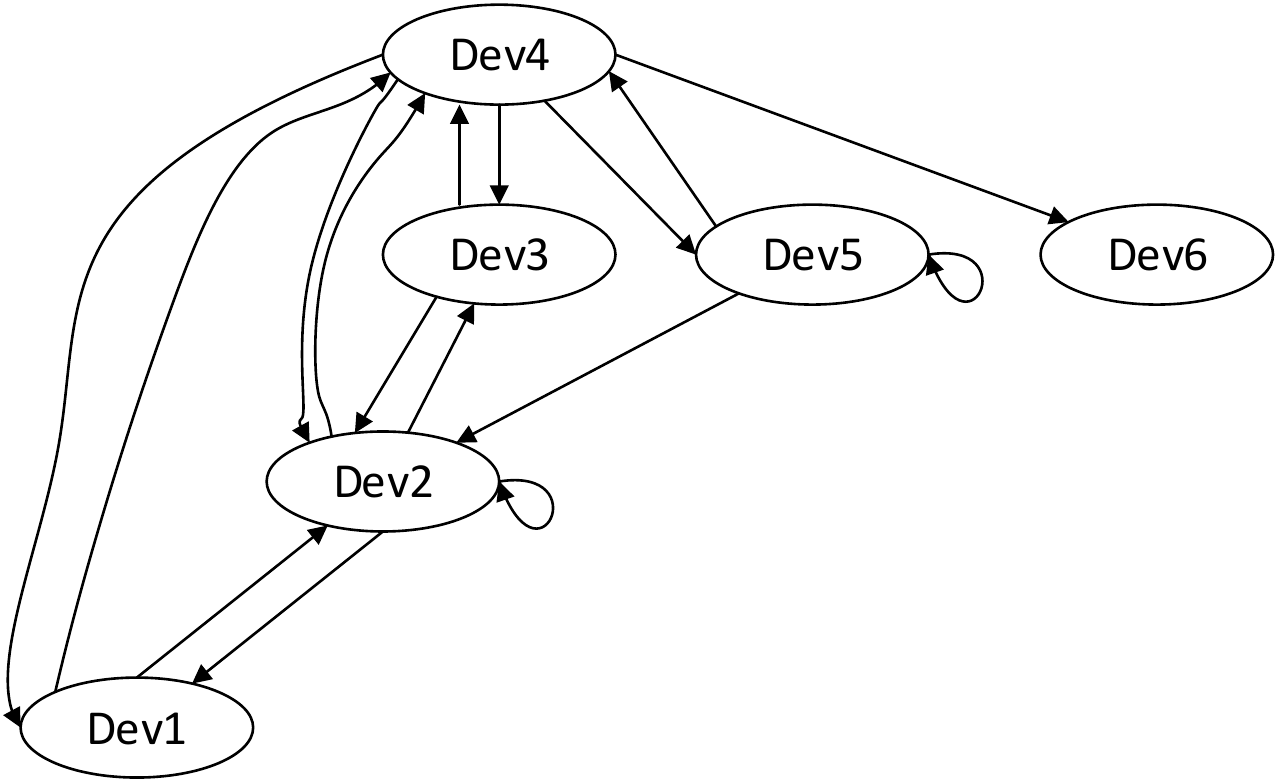}
\caption{Example of a \ac{DSN}. It shows an anonymized excerpt of the \ac{DSN} created by Bird et al.~\protect\citeS{bird2006mining}.}
\label{fig:dsn_example}
\end{figure}

A definition is difficult, because different data sources, research goals, and modelling approaches are used to represent \acp{DSN} in the literature. Due to this, publications on \acp{DSN} contain the specific definition of their \ac{DSN} structure, but this varies between publications. For our purpose, we require a definition, that can be applied to validate if a construct is an instance of a \ac{DSN}. We identified three necessary and sufficient conditions for \acp{DSN}.
\begin{enumerate}
    \item A \ac{DSN} is described by a graph $G = (V,E)$ where $V$ denotes a set of vertices and $E$ a set of edges such that $E \subseteq V \times V$. The graph can be directed or undirected, depending on the intent of the researchers and the data that is used for modelling the \ac{DSN}.
    \item The vertices or a subset of the vertices must represent actors of a software development process, e.g., developers, users, or project managers. 
    \item The edges represent connections between vertices that are based on communication behavior (e.g., email communication) or collaboration behavior (e.g., contributions to the same software artifact). 
\end{enumerate}
An example of a \ac{DSN} is given in Figure~\ref{fig:dsn_example}. This figure depicts an anonymized excerpt of the \ac{DSN} created by Bird et al.~\citeS{bird2006mining}. The vertices in this graph represent different developers, which were active on Apache email lists. A directed edge between two vertices exists, if the developer has sent or replied to at least 150 emails of another developer.



\section{Methodology}
\label{sec:methodology}
Our review follows the guidelines for systematic literature reviews proposed by Kitchenham and Charters~\cite{Kitchenham2007}. Additionally, we used backward and forward snowballing, which was suggested for systematic literature studies by Wohlin~\cite{Wohlin2014}. In the following, we define our underlying research questions, inclusion and exclusion criteria, how we identified papers, and which data was collected for our study. We do not define our study as systematic literature review but as a systematic mapping study, because we did not perform any synthesis of the results, but only provide an overview of the literature. 

\subsection{Research Questions}

In order to study the state of the art in \acp{DSN}, we defined the following five research questions to guide our mapping study. 

\begin{itemize}
  \item \textbf{RQ1.} What software engineering topics have been addressed by \acp{DSN}?
  \item \textbf{RQ2.} Which data sources are used for modelling of \acp{DSN}?
  \item \textbf{RQ3.} What is the scope of the analysis...
  \begin{itemize}
      \item[a)] with respect to number of projects considered
      \item[b)] and people modelled by the \acp{DSN}?
  \end{itemize}
  \item \textbf{RQ4.} What are the most influential...
  \begin{itemize}
      \item[a)] publications?
      \item[b)] authors?
      \item[c)] venues?
  \end{itemize}
  \item \textbf{RQ5.} How did the interest in \ac{DSN} research evolve over time?
\end{itemize}

The first three research questions guide our analysis of the state of the art of \acp{DSN}. We want to get insights into both the topics that are under investigation within the research community, as well as the amount of studies on different topics through our analysis for RQ1. The research questions RQ2 and RQ3 guide our investigation of the scope of studies. Through the answer to RQ2, we want to get valuable information about the data sources that researchers use to define social relationships. Through RQ3, we want to gain insights into how large the studies are, e.g., if they are case studies of specific cases with few projects or if they are broad studies over hundreds of projects. The fourth and fifth question give us insights into the community of \ac{DSN} research itself. RQ4 will tell us which work had the most impact, i.e., early foundational work and later work that presented new ideas for the use of \acp{DSN} that influenced many other publications. Moreover, we assess if there are authors who are clearly distinguished in the field of \ac{DSN} research through their publications. We also look at the venues where \ac{DSN} research is most often published to gain insights into which communities frequently use \acp{DSN} in their research. Through RQ5 we want to understand how the interest in \ac{DSN} research evolves over time, e.g., if the interest is still growing or if the topics of interest change over time.

\subsection{Inclusion and Exclusion Criteria}

To identify which papers should be part of our review, we defined the following
criteria for inclusion:
\begin{itemize}
  \item publications that describe \acp{DSN};
  \item publications that describe how \acp{DSN} may be created; and
  \item publications that describe theoretical aspects of \acp{DSN}.
\end{itemize}

Additionally, we used the following exclusion criteria:
\begin{itemize}
  \item publications that only summarize existing work without new contributions;
  \item publications that only consider social networks or graph structures in general, without a direct and specific relation to software development;
  \item publications that were not peer-reviewed; and
  \item publications that are not published in English.
\end{itemize}

\subsection{Identification of Primary Studies}

\begin{figure}
    \centering
    \includegraphics{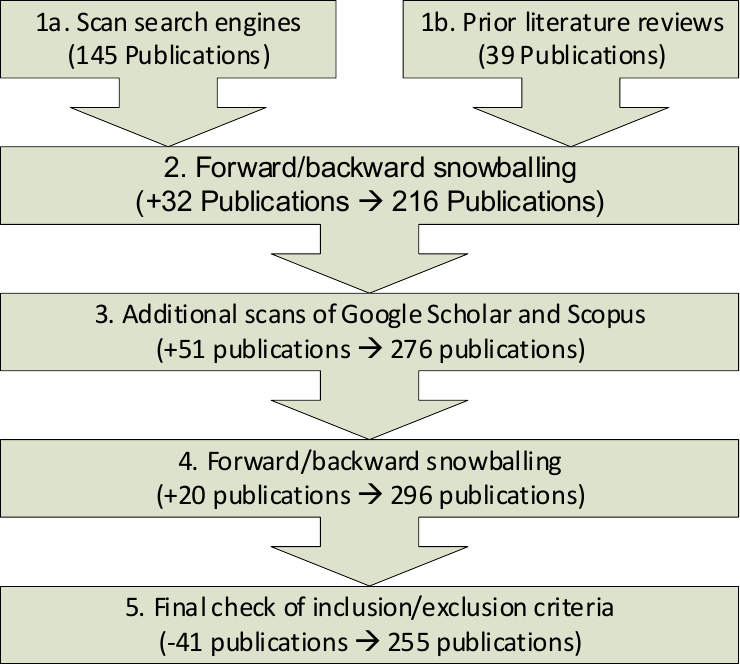}
    \caption{Overview of the mapping study's workflow.}
    \label{fig:workflow}
\end{figure}

Figure~\ref{fig:workflow} summarizes our workflow for the identification of primary studies. We used a five step procedure. 
\begin{enumerate}
    \item Initial scan of the literature using search engines  and prior literature studies to identify a seed of publications. 
    \item Backward and forward snowballing of publications found in the initial scan.
    \item Second scan of the literature using search engines to capture the remainder of 2017 and to account for delayed indexing of publications. 
    \item Backward and forward snowballing of publications found in the second scan.
    \item Final check of inclusion and exclusion criteria on all identified publications.
\end{enumerate}

In the first step, we searched for publications by using five search engines: Google Scholar, IEEE Xplore, ACM Digital Library, Springer Link, Elsevier Search, and Scopus\footnote{Scopus was only used for the additional search in the third step and not for the initial search.}. We used three queries for each search engine: "developer social networks", "developer network", and "collaborative networks OSS". Table~\ref{tbl:searchterms} gives an overview on the number of hits we had with our search terms in each of the search engines. This initial search was conducted between May 2017 and September 2017. Due to the extremely high number of hits, we considered only 750 hits per search engine and search term to get the literature seed for our mapping study. Next, we selected candidates for inclusion by reading the titles, abstracts, and, if it was necessary, the introduction and conclusion sections of the publications. We identified 145 publications through this procedure from the search engines. 
Additionally, we scanned the primary studies from prior related literature studies by Zhang\etal~\cite{zhang2014developer}, Tamburri\etal~\cite{tamburri2013organizational}, Manteli\etal~\cite{manteli2012adopting}, and Abufouda and Abukwaik~\cite{Abufouda2017} (see Section~\ref{sec:relatedwork}). We identified 39 additional publications from the prior studies. This difference is mainly due to the scope of the other literature studies,  especially with respect to search terms. For example, Manteli\etal~\cite{manteli2012adopting} focus on global software engineering and, therefore, also use search terms that do not mention \acp{DSN}. Thus, we identified 184 publications in this first step. 

\begin{table*}
\centering
\begin{tabular}{p{2cm}>{\raggedleft\arraybackslash}p{1.2cm}>{\raggedleft\arraybackslash}p{1.1cm}>{\raggedleft\arraybackslash}p{1.4cm}>{\raggedleft\arraybackslash}p{1.4cm}>{\raggedleft\arraybackslash}p{1.3cm}>{\raggedleft\arraybackslash}p{1.3cm}}
\hline
\textbf{Search terms} & \textbf{Google Scholar} & \textbf{IEEE Xplore} & \textbf{ACM Digital Library} & \textbf{Springer Link} & \textbf{Elsevier Search} & \textbf{Scopus} \\
\hline\hline
developers network & 969,000 & 4,339 & 204,258 & 108,157 & 60,735 & 102\\
developer social networks & 235,000 & 513 & 249,607 & 48,021 & 26,642 & 131  \\
collaborative networks OSS & 25,400 & 22 & 119,424 & 1,090 & 692 & 0 \\
\hline
\end{tabular}
\caption{Search terms and number of hits for each search engine.}
\label{tbl:searchterms}
\end{table*}

In the second step, we checked the related work cited in each of the publications we found using the search engines. This step is also known as backward snowballing~\cite{Wohlin2014}. Moreover, we used the ``cited by'' function of Google Scholar, to identify publications that cited the publications we identified with the search engines. This step is also known as forward snowballing~\cite{Wohlin2014}. We also applied the snowballing to each additional publication we found. We identified 32 additional publications, i.e., 216 publications in total. The snowballing also served to mitigate potential negative effects because we did not consider every hit for the search terms with the search engines. Our assumption is that we find the literature we may have missed through the snowballing. Moreover, same as the use of the prior literature reviews as seed for the snowballing, the snowballing allowed us to identify literature that did not mention the \ac{DSN} in the paper title or abstract and was, therefore, missed by our search. 

In the third step, we repeated our search for literature from the first step. This was required, because the initial search already started in May 2017, i.e., we could not be confident that all papers from 2016 were indexed by the search engines and part of the data for 2017 was not available yet. Moreover, we wanted to include recent publications, that would be missing otherwise. Thus, we repeated the search engines Google Scholar in July 2018 and July 2019 and with SCOPUS in February 2020. This way, we identified 31 new publications using Google Scholar and 29 publications using SCOPUS, bringing our total number of publications to 276. Afterwards, in the fourth step, we performed an additional round of snowballing on these publications and identified 20 additional publications, i.e., a total of 296 publications.

Before we started with the data collection, we validated whether all identified candidates met the inclusion criteria or violate the exclusion criteria in our last step. This way, we excluded 41 of the identified publications, mainly because they were not peer reviewed (e.g., book chapters, preprints on arXiv), summarized only existing work (e.g., surveys, dissertation summaries), or because they did not contain anything specific to developer social networks, regardless of our initial assessment. This left us with 255 primary studies.

\subsection{Data Collection}

Once all literature was identified, we proceeded with the collection of the data required to answer our research questions. For RQ1, we first extracted the research questions and/or hypothesis that were formulated to guide the research, as well as the contributions as listed in the introduction or summarized in the abstract from the publications. We used inductive coding~\cite{Thomas2006} performed by two researchers to identify the research topics of the papers from the hypothesis and contributions in order to obtain the necessary information to answer RQ1. For this, we printed the title, research questions/hypotheses, and contributions of each publication on a separate sheet of paper and sorted them incrementally by their topic, starting with a coarse-grained separation until we were satisfied that our categories provided a sufficient amount of detail for our mapping study. For RQ2 and RQ3, we extracted the data source, the number of projects, and the number of participants in the \ac{DSN} used within the publications. For RQ4 and RQ5, we collected meta data about the publications themselves, i.e., the title, authors, publication venue, year, and number of citations. We organized the collected data in a spreadsheet which is made available as supplementary material. 

\section{Literature Review}
\label{sec:review}
In this section, we provide the review of the the state of the art of \ac{DSN} research based on the data collection we described in Section~\ref{sec:methodology}. We systematically address different topics. We use the data from this review to answer our research questions in Section~\ref{sec:discussion}.

\subsection{Research Directions}
\label{sec:research-directions}

Based on the description of the contributions,  the research questions, and the research hypotheses of publications, we identified seven general research directions regarding \acp{DSN}. For four of the general research directions we identified subtopics, i.e., specific aspects that were considered within the general direction. Table~\ref{tbl:researchdirections} shows our mapping of publications to the research directions including subtopics. 

Nearly half of the publications we identified analyze the \textbf{community structures} in software development projects. Most of these publications analyzed the general structure of the \ac{DSN}. However, we also identified seven more specific subtopics of the analysis of community structures: the evolution of the communities by considering \acp{DSN} over time; community structures in the context of global software engineering; the formation of teams within development projects; the correlation between the community structure and code quality; the analysis of socio-technical congruence; the simulation of community structures; and the identification of community smells. 

\acp{DSN} are frequently used for the creation or improvement of \textbf{prediction} models for various aspects in software development projects. We identified seven subtopics of prediction approaches using \acp{DSN}: bug triage, i.e., support for assigning appropriate developers to work on bug reports; defect prediction, i.e., using the social structure of a project to enhance models that estimate the defect-proneness of different parts of software; recommendation of suitable developers for project work in general; predictions of the outcome of a project, i.e., if projects are likely successful; predictions of suitable Web services; predictions of build failures; and prediction of appropriate developers for code review. 

The \textbf{collaboration behavior} was also scrutinized using \acp{DSN}. While \acp{DSN} are modelling some direct or indirect collaboration behavior in software development projects, the analysis of the collaboration behavior itself is in general not the focus. The publications we identified for this research direction focus directly on the collaboration behavior, e.g., which tools were used or how collaboration behavior was impacted by the structure of projects. In addition to research on collaboration behavior in general, we identified three more specific subtopics: collaboration behavior in global software engineering; problems in collaboration behavior and how they are reflected in \acp{DSN}; and collaboration between developers from different companies, including competitors in open source projects. 

\acp{DSN} are also frequently used to assess the \textbf{roles of developers} within a development project, e.g., whether a developer is a core developer or a peripheral developer. While the identification of roles for developers in general is the main topic of this research direction, we also identified two other subtopics; the analysis of how onboarding of peripheral developers within projects works; and how developers specialize within a project. 

We also identified research regarding \textbf{tools} for \ac{DSN} analysis, mostly for the visualization of \acp{DSN} based on different information sources. 

The \textbf{validity of \ac{DSN} research} was also considered by five publications. These publications do not question the validity of \ac{DSN} research in general, but rather analyze how properties of \ac{DSN} research may depend on the specific context of research projects, e.g., the scope of the analysis or the repository that was used as source for the \acp{DSN}. 

Finally, we found one publication on a \textbf{data set} that directly contains the graph structure of a \ac{DSN}. The lack of publications on data sets shows that researchers either generate \acp{DSN} from data they collect, or from more general data sets that do not model \acp{DSN} directly. Such data sets contain general information mined from software repositories from which a \ac{DSN} is then built.

\begin{table}
\footnotesize
\begin{tabular}{p{3.7cm}>{\raggedleft\arraybackslash}p{1.4cm}p{5.6cm}}
\hline
\textbf{Category} & \textbf{\#Pubs.} & \textbf{Publications} \\
\hline\hline
\multicolumn{3}{l}{\textbf{Community Structure}} \\
General & 75 & \citeS{allaho2013analyzing,amrit2004social,antwerp2010importance,avelino2019measuring,bana2018influence,batista2017collaboration,behfar2016intragroup,behfar2018knowledge,bidoki2018network,bird2006mining,bird2006mining1,bird2008latent,canfora2011social,cherry2008social,conaldi2010meso,crowston2005social,crowston2006hierarchy,dos2011bringing,gao2007network,gao2007towards,geipel2014communication,gloor2003visualization,gonzalez2004community,guilherme2017assessing,he2012applying,howison2006social,hu2008comparison,hu2012reputation,hu2016influence,huang2011relating,ichimura2015analysis,iyer2019requirements,jermakovics2011mining,jermakovics2013exploring,jiang2013understanding,joblin2015developer,kamei2008analysis,kidane2007correlating,leibzon2016social,lim2011evolving,lima2014coding,linaaker2019method,long2007social,lopez2004applying,lopez2008applying,madey2002open,meneely2009secure,meneely2010strengthening,meneely2010use,mergel2015open,nzeko2015social,qiu2019going,robertsa2006communication,schwind2008unveiling,singh2010small,sowe2014empirical,sureka2011using,surian2010mining,tamburri2019discovering,tan2007social,thung2013network,toral2010analysis,van2010importance,wagstrom2005social,wang2019investigating,wiggins2008social,wolf2009mining,xu2004exploration,xu2005open,xu2005topological,yu2013study,yu2014exploring,zanetti2012quantitative,zhang2014generative,zhang2015analyzing} \\
DSN Evolution & 18 & \citeS{aljemabi2018empirical,datta2011evolution,hannemann2013community,hong2011understanding,joblin2017evolutionary,kakimoto2006social,kavaler2017stochastic,kumar2013evolution,kumar2019studying,nakakoji2005understanding,ngamkajornwiwat2008exploratory,ryan2010modeling,sharma2011studying,van2010open,weiss2006evolution,yu2014exploring,zanetti2013rise,zhang2011network} \\
Global SWE & 10 & \citeS{ahuja2003individual,avritzer2010coordination,cataldo2008communication1,de2007toward,ehrlich2006leveraging,ehrlich2012all,hinds2006structures,hossain2009social,sarker2011path,spinellis2006global} \\
Team Formation & 6 & \citeS{caglayan2013emergence,crowston2007self,hahn2006impact,hahn2008emergence,panichella2014evolution,singh2010developer} \\
Impact on Code Quality & 6 & \citeS{bettenburg2010studying,bettenburg2013studying,ccaglayan2016effect,datta2018does,hossain2008measuring,mockus2010organizational} \\
Socio-technical Congruence & 5 & \citeS{cataldo2008socio,de2005seeking,kwan2011does,syeed2013socio,valetto2007using} \\
Simulation & 4 & \citeS{honsel2014software,honsel2015developer,honsel2015mining,yu2008mining} \\
Community Smells & 2 & \citeS{catolino2019gender,palomba2018beyond} \\
\hline
\multicolumn{3}{l}{\textbf{Prediction}} \\
Bug Triage & 16 & \citeS{banitaan2013decoba,bhattacharya2010fine,bhattacharya2012automated,chen2010improving,jeong2009improving,sun2017enhancing,wang2013devnet,wu2011drex,wu2018empirical,xuan2012developer,yang2014utilizing,zanetti2013categorizing,zhang2012automated,zhang2013heterogeneous,zhang2014butter,zhang2014novel} \\
Defect Prediction & 12 & \citeS{abreu2009developer,bhattacharya2012graph,biccer2011defect,bird2009putting,hu2013using,meneely2008predicting,miranskyy2014effect,pinzger2008can,simpson2013changeset,wang2016analyzing,zhang2014mining,zhang2019file} \\
Project Outcomes & 9 & \citeS{cataldo2012impact,jarczyk2018surgical,liu2007design,peng2018co,peng2019co,singh2011network,surian2013predicting,wang2012human,wang2012survival} \\
Developers for Tasks in General & 7 & \citeS{dravzdilova2012method,hossain2006actor,hu2018user,li2016task,mcdonald2003recommending,surian2011recommending,wan2018scsminer} \\
Suitable Web Services & 3 & \citeS{bianchini2015developers,bianchini2016role,bianchini2016social} \\
Build Failures & 2 & \citeS{schroter2010predicting,wolf2009predicting} \\
Developers for Code Review & 1 & \citeS{kerzazi2016can} \\
\hline
\multicolumn{3}{l}{\textbf{Collaboration Behavior}} \\
General & 13 & \citeS{cohen2018large,damian2013role,duc2011impact,feczak2009measuring,feczak2011exploring,gharehyazie2017tracing,kerzazi2017knowledge,licorish2017exploring,omoronyia2009using,ortu2015measuring,wu2016effects,xuan2012measuring,yang2014social} \\
Global SWE & 10 & \citeS{cataldo2008communication,chang2007out,damian2007collaboration,fonseca2006exploring,herbsleb2003empirical,mikawa2009removing,nguyen2008global,sarker2011role,urdangarin2008experiences,wolf2008does} \\
Problems & 10 & \citeS{begel2010codebook,bernardi2012developers,bhowmik2016optimal,cataldo2006identification,damian2007awareness,ell2013identifying,orsila2009trust,Sapkota2020,wang2016diffusion,xuan2016converging} \\
Inter-company collaboration behavior & 1 & \citeS{teixeira2015lessons} \\
\hline
\multicolumn{3}{l}{\textbf{Developer Roles}} \\
Identification & 18 & \citeS{zhang2012empirical,meneely2010improving,bhattacharya2014determining,crowston2006core,datta2010social,dittrich2013network,huang2005mining,joblin2017classifying,lee2013github,licorish2014understanding,licorish2015communication,lim2010stakenet,lim2012stakerare,marczak2008information,pohl2008dynamic,sharma2017boundary,sowe2006identifying,yu2007mining} \\
Onboarding & 9 & \citeS{bird2007open,canfora2012who,casalnuovo2015developer,cheng2017developer,ducheneaut2005socialization,el2017periphery,gharehyazie2013social,gharehyazie2015developer,zhou2011does} \\
Specialization & 1 & \citeS{maclean2011knowledge} \\
\hline
\textbf{Tools} & 11 & \citeS{borici2012proxiscientia,de2004technical,de2007supporting,gilbert2007codesaw,gote2019git,ogawa2007visualizing,ohira2005accelerating,ohira2005supporting,sarma2009tesseract,schwind2008svnnat,schwind2010tool} \\
\hline
\textbf{DSN Validity} & 5 & \citeS{meneely2011socio,aljemabi2017empirical,nia2010validity,panichella2014developers,tymchuk2014collaboration} \\
\hline
\textbf{Datasets} & 1 & \citeS{maclean2013apache} \\
\hline
\end{tabular}
\caption{Overview of the literature on \acp{DSN} by research directions.}
\label{tbl:researchdirections}
\end{table}

\begin{mdframed}
\textbf{Answer to RQ\,1:} 
Community structures are the dominant research direction. Other frequently studied directions are \acp{DSN} for predictions, collaboration behavior and developer roles. Tools, studies on validity, and data sets play only a minor role. 
\end{mdframed}

\subsection{Data Sources}

There are five major data sources which are used by 241 of the 255 publications:
\begin{itemize}
    \item Forges like GitHub or SourceForge that are used by millions of developers for hosting and developing open source software. These forges offer an integration of \acp{VCS} and \acp{ITS} within a single environment, often coupled with other services like Web pages, hosting of releases, or Wikis. Thus, they are a rich source for collaborations between developers, both within a project, as well as across multiple projects.
    \item \acp{ITS} like Jira or Bugzilla are used for the collection, tracking, and management of issues and work items within projects, e.g., change requests, bug reports, or questions by users. \acp{ITS} allow the discussion about issues, the definition of work flows for issues, and different types of resolutions.
    \item \acp{VCS} like Git or SVN are systems that track and archive changes of files and folders over time. Typically, \acp{VCS} allow different development branches and support working collaboratively on the same resources~\cite{sommerville2011software}. 
    \item \acp{ML} are collections of email addresses that can be used for communication within software projects. \acp{ML} may be restricted, e.g., not everybody may be allowed to post or subscribe to a \ac{ML}. Participants of \acp{ML} may be natural persons (e.g., developers, users), but also systems (e.g., continuous integration systems, \acp{ITS}).
    \item Surveys, i.e., interviews or questionnaires that were used to directly ask developers about their communication behaviour within a development project. 
\end{itemize}

In addition to the five major sources, there are other ways that researchers used to collect information about collaboration behavior which we summarized as "Other" in Table~\ref{tbl:datasources}. These are IRC chats~\citeS{cataldo2008communication1,cataldo2008communication,panichella2014developers,wang2016diffusion}, plug-ins that monitor development environments~\citeS{omoronyia2009using,borici2012proxiscientia,de2007supporting}, manual inspection of project documents, e.g., requirements~\citeS{damian2013role,damian2007collaboration,marczak2008information}, owners ob web service mash-ups~\citeS{bianchini2016role, bianchini2016social, bianchini2015developers}, 
the web site Ohloh that provides statistics about open source development\footnote{The name has changed to https://www.openhub.net/.}~\citeS{hu2008comparison,hu2012reputation}, online discussion forums~\citeS{wiggins2008social,crowston2007self}, JAR files~\citeS{hu2013using}, the BlogLinks and Advogoto social networks\footnote{Both are not available online anymore.} of software developers~\citeS{wagstrom2005social}, on site researchers that observe communication behavior~\citeS{damian2007awareness}, employee directories~\citeS{begel2010codebook}, and the code review portal Gerrit~\citeS{yang2014social}. Additionally, one publication discusses \acp{DSN} from an abstract perspective and proposes the use of tracking for every communication including phone calls, emails, etc.~\citeS{amrit2004social}.

\begin{figure}
    \centering
    \includegraphics[width=0.8\textwidth]{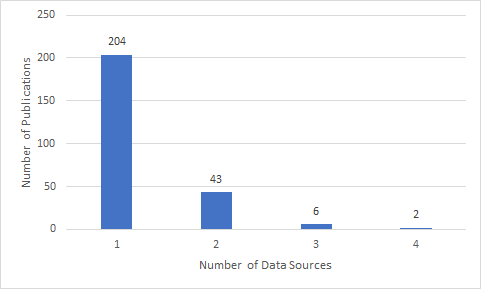}
    \caption{Number of data sources that were used for the modelling of the \acp{DSN} within the publications.}
    \label{fig:num_data_sources_in_pubs}
\end{figure}

Figure~\ref{fig:num_data_sources_in_pubs} depicts the number of data sources that were used for modelling \acp{DSN}. It highlights that 204 of the 255 publications build a \ac{DSN} that is based on a single source, 43 publications used a combination of two data sources, six publication three data sources and two publications four data sources.

\begin{table}
\footnotesize
\begin{tabular}{p{4cm}>{\raggedleft\arraybackslash}p{1.2cm}p{5.6cm}}
\hline
\textbf{Data Source} & \textbf{\#Pubs.} & \textbf{Publications} \\
\hline\hline
Forge & 64 & \citeS{aljemabi2018empirical,allaho2013analyzing,bana2018influence,batista2017collaboration,behfar2016intragroup,behfar2018knowledge,bidoki2018network,caglayan2013emergence,catolino2019gender,cohen2018large,conaldi2010meso,dos2011bringing,dravzdilova2012method,el2017periphery,gao2007network,gao2007towards,hahn2006impact,hahn2008emergence,he2012applying,hu2016influence,hu2018user,huang2011relating,ichimura2015analysis,iyer2019requirements,jarczyk2018surgical,jiang2013understanding,kerzazi2016can,kerzazi2017knowledge,lee2013github,leibzon2016social,li2016task,lima2014coding,liu2007design,madey2002open,mergel2015open,ohira2005accelerating,ohira2005supporting,peng2018co,peng2019co,qiu2019going,Sapkota2020,singh2010small,singh2011network,surian2010mining,surian2011recommending,surian2013predicting,tamburri2019discovering,tan2007social,thung2013network,tymchuk2014collaboration,van2010importance,wan2018scsminer,wang2012human,wang2012survival,wang2019investigating,wu2016effects,xu2004exploration,xu2005open,xu2005topological,yu2013study,yu2014exploring,yu2014exploring,zhang2014generative,zhang2015analyzing} \\
ITS & 49 & \citeS{abreu2009developer,banitaan2013decoba,bettenburg2010studying,bhattacharya2010fine,bhattacharya2012automated,bhowmik2016optimal,cataldo2006identification,cataldo2008socio,cataldo2012impact,chen2010improving,crowston2005social,crowston2006core,crowston2006hierarchy,datta2010social,duc2011impact,ehrlich2012all,feczak2009measuring,feczak2011exploring,hong2011understanding,hossain2008measuring,hossain2009social,howison2006social,jeong2009improving,kumar2013evolution,kumar2019studying,licorish2014understanding,licorish2015communication,licorish2017exploring,long2007social,nguyen2008global,ortu2015measuring,sharma2011studying,sureka2011using,wang2013devnet,wolf2008does,wolf2009mining,wolf2009predicting,wu2011drex,wu2018empirical,xuan2012developer,yang2014utilizing,zanetti2012quantitative,zanetti2013categorizing,zanetti2013rise,zhang2012automated,zhang2013heterogeneous,zhang2014butter,zhang2014novel,zhou2011does} \\
VCS & 41 & \citeS{antwerp2010importance,avelino2019measuring,bird2009putting,casalnuovo2015developer,ccaglayan2016effect,cheng2017developer,de2004technical,de2005seeking,dittrich2013network,ell2013identifying,gonzalez2004community,gote2019git,guilherme2017assessing,huang2005mining,jermakovics2011mining,jermakovics2013exploring,joblin2015developer,joblin2017classifying,joblin2017evolutionary,kakimoto2006social,lopez2004applying,lopez2008applying,maclean2013apache,meneely2008predicting,meneely2009secure,meneely2010strengthening,meneely2011socio,miranskyy2014effect,mockus2010organizational,orsila2009trust,palomba2018beyond,pinzger2008can,pohl2008dynamic,schwind2008svnnat,schwind2008unveiling,schwind2010tool,sun2017enhancing,teixeira2015lessons,valetto2007using,van2010open,yu2007mining} \\
ML & 23 & \citeS{ahuja2003individual,bird2006mining,bird2006mining1,bird2008latent,gharehyazie2015developer,gloor2003visualization,hossain2006actor,kamei2008analysis,kavaler2017stochastic,kidane2007correlating,nakakoji2005understanding,ngamkajornwiwat2008exploratory,nia2010validity,nzeko2015social,robertsa2006communication,sharma2017boundary,sowe2006identifying,toral2010analysis,weiss2006evolution,xuan2016converging,yu2008mining,zhang2012empirical,zhang2014mining} \\
Other & 14 & \citeS{amrit2004social,bianchini2015developers,bianchini2016role,bianchini2016social,borici2012proxiscientia,cataldo2008communication,damian2007awareness,de2007supporting,hu2008comparison,hu2012reputation,hu2013using,omoronyia2009using,wang2016analyzing,yang2014social} \\ 
Survey & 13 & \citeS{avritzer2010coordination,chang2007out,cherry2008social,ehrlich2006leveraging,hinds2006structures,lim2010stakenet,lim2011evolving,lim2012stakerare,mcdonald2003recommending,mikawa2009removing,sarker2011path,sarker2011role,urdangarin2008experiences} \\ \hdashline
ITS \& VCS & 19 & \citeS{aljemabi2017empirical,bernardi2012developers,bettenburg2013studying,bhattacharya2012graph,bhattacharya2014determining,biccer2011defect,datta2011evolution,datta2018does,honsel2014software,honsel2015developer,honsel2015mining,kwan2011does,linaaker2019method,meneely2010improving,meneely2010use,schroter2010predicting,simpson2013changeset,spinellis2006global,zhang2019file} \\
ML \& VCS & 14 & \citeS{bird2007open,canfora2011social,canfora2012who,ducheneaut2005socialization,fonseca2006exploring,gharehyazie2017tracing,gilbert2007codesaw,hannemann2013community,ogawa2007visualizing,singh2010developer,sowe2014empirical,syeed2013socio,xuan2012measuring,zhang2011network} \\
Survey \& Other & 3 & \citeS{damian2007collaboration,damian2013role,marczak2008information} \\
ML \& ITS & 2 & \citeS{maclean2011knowledge,ryan2010modeling} \\
ML \& Other & 2 & \citeS{crowston2007self,wagstrom2005social} \\
Forge \& Survey & 1 & \citeS{de2007toward} \\
ITS \& Survey & 1 & \citeS{herbsleb2003empirical} \\
VCS \& Forge & 1 & \citeS{geipel2014communication} \\ \hdashline
ITS, ML \& VCS & 3 & \citeS{gharehyazie2013social,panichella2014evolution,sarma2009tesseract} \\
ML, ITS \& Other & 1 & \citeS{wiggins2008social} \\
ITS, Survey \& Other & 1 & \citeS{cataldo2008communication1} \\
ITS, VCS \& Other & 1 & \citeS{begel2010codebook} \\ \hdashline
ITS, ML, CVS \& Other & 2 & \citeS{panichella2014developers,wang2016diffusion} \\
\hline
\end{tabular}
\caption{Data sources that were used for the modelling of the \acp{DSN}.}
\label{tbl:datasources}
\end{table}

\begin{mdframed}
\textbf{Answer to RQ\,2:}
Software repositories like forges, \acp{ITS}, \acp{VCS} and \acp{ML} are the main sources for \acp{DSN}, however, surveys are also sometimes used. Publications commonly use a single source for \ac{DSN} modelling. The knowledge about \acp{DSN} built with multiple sources is limited.  
\end{mdframed}

\subsection{Number of Projects Analyzed}
\label{sec:num_projects}

A major factor regarding the external validity of results is the number of projects for which data is collected. If only data about very few projects is used for an empirical study about a phenomenon that can be studied using \acp{DSN}, the results may not generalize to other projects. The likelihood that the results generalize to software engineering in general increases with the number of projects that are analyzed. Table~\ref{tbl:numprojects} shows the number of projects per publication. The data we collected shows that most papers on \acp{DSN} perform some sort of empirical study to demonstrate their approach or research a phenomenon. Only 12 of the 255 publications we identified did not perform any empirical study. Moreover, we identified 16 publications for which we could not identify the number of projects from the publication. There were two reasons for this: either the authors did not report how they selected a smaller subset from a larger database or the authors did not specify which projects were used at all. This is not only problematic for evaluating the external validity of a study, but also hinders replications of the results. Of the 227 publications for which we could identify the number of projects, 76 used only a single project for their empirical study, 69 used only 2-5 projects for the empirical study. In other words, about 33\% of the publications on DSNs used a single project, another 30\% used 2-5 projects. Both numbers are extremely low and do not allow for a generalization of the findings due to the limited context covered by the projects. Another 12 publications only considered 6-10 projects, which is still a small number. On the bright side, 50 publications used more than 100 projects, i.e., larger sample sizes that usually allow to generalize findings. 38 of these publications use a forge as data source. Regardless, our analysis of the sample sizes with respect to the number of projects indicates a severe threat to the external validity of many empirical studies on \acp{DSN}.

\begin{table}
\begin{tabular}{p{1.6cm}>{\raggedleft\arraybackslash}p{1.2cm}p{8cm}}
\hline
\textbf{\#Projects} & \textbf{\#Pubs.} & \textbf{Publications} \\
\hline\hline
1 & 76 & \citeS{abreu2009developer,ahuja2003individual,avritzer2010coordination,bettenburg2010studying,bird2006mining,bird2006mining1,caglayan2013emergence,cataldo2006identification,cataldo2008communication,cataldo2008socio,cataldo2012impact,cherry2008social,damian2007awareness,datta2010social,datta2011evolution,datta2018does,ducheneaut2005socialization,ehrlich2012all,ell2013identifying,gonzalez2004community,he2012applying,hong2011understanding,honsel2014software,honsel2015developer,hossain2006actor,hu2008comparison,kamei2008analysis,kumar2013evolution,kwan2011does,li2016task,licorish2014understanding,licorish2015communication,licorish2017exploring,lim2010stakenet,lim2011evolving,lim2012stakerare,linaaker2019method,maclean2011knowledge,marczak2008information,mcdonald2003recommending,meneely2008predicting,meneely2009secure,meneely2010improving,meneely2010use,mikawa2009removing,miranskyy2014effect,mockus2010organizational,nakakoji2005understanding,ngamkajornwiwat2008exploratory,nguyen2008global,omoronyia2009using,orsila2009trust,pinzger2008can,pohl2008dynamic,robertsa2006communication,ryan2010modeling,sarma2009tesseract,schroter2010predicting,sharma2011studying,sharma2017boundary,simpson2013changeset,spinellis2006global,sureka2011using,syeed2013socio,toral2010analysis,urdangarin2008experiences,wolf2008does,wolf2009mining,wolf2009predicting,wu2011drex,yang2014utilizing,zanetti2013rise,zhang2011network,zhang2012automated,zhang2012empirical,zhang2014butter} \\
2-5 & 69 & \citeS{aljemabi2018empirical,banitaan2013decoba,bernardi2012developers,bettenburg2013studying,bhattacharya2010fine,bhattacharya2012automated,bhattacharya2014determining,bhowmik2016optimal,biccer2011defect,bird2007open,bird2008latent,bird2009putting,canfora2011social,canfora2012who,cataldo2008communication1,ccaglayan2016effect,chang2007out,chen2010improving,crowston2007self,damian2013role,de2007supporting,dittrich2013network,duc2011impact,ehrlich2006leveraging,el2017periphery,gilbert2007codesaw,gloor2003visualization,hannemann2013community,honsel2015mining,hu2013using,jeong2009improving,jermakovics2011mining,jermakovics2013exploring,kakimoto2006social,kavaler2017stochastic,kerzazi2016can,kerzazi2017knowledge,kidane2007correlating,kumar2019studying,leibzon2016social,lopez2004applying,lopez2008applying,meneely2010strengthening,meneely2011socio,nia2010validity,nzeko2015social,ogawa2007visualizing,panichella2014evolution,sarker2011role,schwind2008svnnat,schwind2008unveiling,singh2010developer,sowe2006identifying,sun2017enhancing,van2010open,wang2013devnet,wang2016analyzing,wang2016diffusion,wiggins2008social,wu2018empirical,xuan2012developer,yang2014social,yu2007mining,yu2008mining,zanetti2013categorizing,zhang2013heterogeneous,zhang2014mining,zhang2014novel,zhang2019file} \\
6-10 & 12 & \citeS{bhattacharya2012graph,gharehyazie2013social,gharehyazie2015developer,guilherme2017assessing,huang2005mining,joblin2015developer,joblin2017classifying,ortu2015measuring,palomba2018beyond,panichella2014developers,teixeira2015lessons,zhou2011does} \\
11-100 & 20 & \citeS{aljemabi2017empirical,catolino2019gender,crowston2005social,de2007toward,dos2011bringing,geipel2014communication,gharehyazie2017tracing,hinds2006structures,hossain2008measuring,hossain2009social,joblin2017evolutionary,sarker2011path,sowe2014empirical,tamburri2019discovering,weiss2006evolution,xuan2012measuring,xuan2016converging,zanetti2012quantitative,zhang2014generative,zhang2015analyzing} \\
$>$100 & 50 & \citeS{allaho2013analyzing,antwerp2010importance,avelino2019measuring,batista2017collaboration,bianchini2015developers,bianchini2016role,bidoki2018network,casalnuovo2015developer,cheng2017developer,cohen2018large,conaldi2010meso,crowston2006core,crowston2006hierarchy,dravzdilova2012method,feczak2009measuring,feczak2011exploring,hahn2006impact,hahn2008emergence,howison2006social,hu2012reputation,hu2018user,huang2011relating,iyer2019requirements,jarczyk2018surgical,jiang2013understanding,lee2013github,liu2007design,long2007social,madey2002open,mergel2015open,ohira2005accelerating,peng2018co,peng2019co,Sapkota2020,singh2010small,singh2011network,surian2010mining,surian2011recommending,surian2013predicting,tan2007social,thung2013network,tymchuk2014collaboration,wan2018scsminer,wang2012human,wang2012survival,wang2019investigating,wu2016effects,xu2005open,xu2005topological,yu2013study} \\
Missing & 16 & \citeS{bana2018influence,behfar2016intragroup,behfar2018knowledge,bianchini2016social,damian2007collaboration,gao2007network,herbsleb2003empirical,hu2016influence,lima2014coding,maclean2013apache,qiu2019going,van2010importance,wagstrom2005social,xu2004exploration,yu2014exploring,yu2014exploring} \\
NA & 12 & \citeS{amrit2004social,begel2010codebook,borici2012proxiscientia,de2004technical,de2005seeking,fonseca2006exploring,gao2007towards,gote2019git,ichimura2015analysis,ohira2005supporting,schwind2010tool,valetto2007using} \\
\hline
\end{tabular}
\caption{Number of projects that were analyzed as part of an empirical study of \acp{DSN}. Missing means that the number of projects is not or not accurately reported in the publication, NA means that the publication did not conduct an empirical study.}
\label{tbl:numprojects}
\end{table}

\begin{mdframed}
\textbf{Answer to RQ\,3a:}
Over 69\% of all publications use less than 11 projects to evaluate their findings. Most publications with at least 100 projects use a forge as data source (38 of 50). 
\end{mdframed}

\subsection{Number of Developers in the \acp{DSN}}

The second major factor regarding the validity of results is the number of people that are part of the \acp{DSN}. Table~\ref{tbl:numdevs} shows the data we collected regarding the number of people in the \acp{DSN}. In case a publication created multiple \acp{DSN}, e.g., one per project considered, we report the mean value of the people in the \acp{DSN}. The number of people modelled by the \acp{DSN} is relatively high. 77 publications have more than 1,000 people as part of their \acp{DSN}, 15 publications actually model more than 100,000 people. Only four publications have very small networks with less than or equal to 10 people, another 32 publications consider less than or equal to 100 people. Thus, for the publications for which the data about the number of people is available, the networks that are considered are in general relatively large. When we looked closely at the data, we observed two reasons for this: first, while many publications consider only few projects, these projects tend to be very large, e.g., Mozilla Firefox and the Eclipse IDE. Moreover, our data also shows that \acp{ML} and forges are the most common data sources for \acp{DSN}. Both capture not only developers, but also users of the respective projects. We also found a very concerning general trend in the literature: 66 of the 240 publications that performed an empirical study did not report the number of participants in the \ac{DSN}. This is a vital piece of information for the estimation of both the internal and external validity of empirical studies that should always be reported. 

\begin{table}
\begin{tabular}{p{1.6cm}>{\raggedleft\arraybackslash}p{1.2cm}p{8cm}}
\hline
\textbf{\#People} & \textbf{\#Pubs.} & \textbf{Publications} \\
\hline\hline
1-10 & 4 & \citeS{cherry2008social,damian2007collaboration,omoronyia2009using,pohl2008dynamic} \\
11-100 & 32 & \citeS{abreu2009developer,ahuja2003individual,avritzer2010coordination,banitaan2013decoba,bird2008latent,cataldo2012impact,chang2007out,crowston2007self,damian2013role,datta2010social,de2007toward,ehrlich2006leveraging,gharehyazie2017tracing,honsel2015mining,huang2005mining,jarczyk2018surgical,jermakovics2011mining,kakimoto2006social,kavaler2017stochastic,lim2011evolving,marczak2008information,meneely2010improving,meneely2010use,mikawa2009removing,palomba2018beyond,panichella2014developers,sarker2011path,sarker2011role,schwind2008unveiling,urdangarin2008experiences,yu2007mining,zhang2012empirical} \\
101-1000 & 64 & \citeS{aljemabi2017empirical,bianchini2015developers,bianchini2016role,caglayan2013emergence,canfora2012who,cataldo2006identification,cataldo2008communication,cataldo2008socio,crowston2006hierarchy,datta2011evolution,datta2018does,dittrich2013network,duc2011impact,ducheneaut2005socialization,ehrlich2012all,gharehyazie2013social,gharehyazie2015developer,gloor2003visualization,guilherme2017assessing,herbsleb2003empirical,honsel2015developer,hossain2006actor,huang2011relating,jermakovics2013exploring,joblin2015developer,joblin2017classifying,joblin2017evolutionary,kamei2008analysis,kerzazi2017knowledge,kwan2011does,licorish2014understanding,licorish2015communication,licorish2017exploring,lim2012stakerare,linaaker2019method,lopez2008applying,maclean2011knowledge,meneely2008predicting,meneely2009secure,meneely2010strengthening,meneely2011socio,mockus2010organizational,ngamkajornwiwat2008exploratory,nguyen2008global,orsila2009trust,ortu2015measuring,robertsa2006communication,ryan2010modeling,schwind2008svnnat,sowe2014empirical,spinellis2006global,sun2017enhancing,surian2011recommending,tamburri2019discovering,tymchuk2014collaboration,weiss2006evolution,wolf2008does,wolf2009predicting,wu2011drex,xuan2012measuring,zanetti2012quantitative,zanetti2013categorizing,zhang2011network,zhang2012automated} \\
1001-10000 & 35 & \citeS{aljemabi2018empirical,avelino2019measuring,behfar2016intragroup,behfar2018knowledge,bhowmik2016optimal,bidoki2018network,bird2006mining,bird2006mining1,canfora2011social,casalnuovo2015developer,el2017periphery,hannemann2013community,he2012applying,honsel2014software,hu2008comparison,jeong2009improving,kerzazi2016can,kumar2019studying,leibzon2016social,li2016task,lim2010stakenet,liu2007design,nakakoji2005understanding,nia2010validity,nzeko2015social,ogawa2007visualizing,singh2010small,sowe2006identifying,sureka2011using,syeed2013socio,toral2010analysis,xuan2016converging,yu2008mining,zhang2014butter,zhang2019file} \\
10001-100000 & 27 & \citeS{antwerp2010importance,batista2017collaboration,bhattacharya2014determining,bird2007open,cohen2018large,dravzdilova2012method,hu2012reputation,kumar2013evolution,long2007social,madey2002open,panichella2014evolution,qiu2019going,Sapkota2020,sarma2009tesseract,sharma2017boundary,surian2010mining,tan2007social,thung2013network,van2010importance,van2010open,wan2018scsminer,wang2013devnet,wu2018empirical,xuan2012developer,zanetti2013rise,zhang2013heterogeneous,zhou2011does} \\
$>$100000 & 15 & \citeS{allaho2013analyzing,bernardi2012developers,conaldi2010meso,gao2007network,hahn2008emergence,hong2011understanding,hu2018user,jiang2013understanding,lima2014coding,ohira2005accelerating,wang2019investigating,xu2005topological,yu2013study,yu2014exploring,yu2014exploring} \\
Missing & 66 & \citeS{bana2018influence,bettenburg2010studying,bettenburg2013studying,bhattacharya2010fine,bhattacharya2012automated,bhattacharya2012graph,bianchini2016social,biccer2011defect,bird2009putting,cataldo2008communication1,catolino2019gender,ccaglayan2016effect,chen2010improving,cheng2017developer,crowston2005social,crowston2006core,damian2007awareness,de2007supporting,dos2011bringing,ell2013identifying,feczak2009measuring,feczak2011exploring,geipel2014communication,gilbert2007codesaw,gonzalez2004community,hahn2006impact,hinds2006structures,hossain2008measuring,hossain2009social,howison2006social,hu2013using,hu2016influence,iyer2019requirements,kidane2007correlating,lee2013github,lopez2004applying,maclean2013apache,mcdonald2003recommending,mergel2015open,miranskyy2014effect,peng2018co,peng2019co,pinzger2008can,schroter2010predicting,sharma2011studying,simpson2013changeset,singh2010developer,singh2011network,surian2013predicting,teixeira2015lessons,wagstrom2005social,wang2012human,wang2012survival,wang2016analyzing,wang2016diffusion,wiggins2008social,wolf2009mining,wu2016effects,xu2004exploration,xu2005open,yang2014social,yang2014utilizing,zhang2014generative,zhang2014mining,zhang2014novel,zhang2015analyzing} \\
NA & 12 & \citeS{amrit2004social,begel2010codebook,borici2012proxiscientia,de2004technical,de2005seeking,fonseca2006exploring,gao2007towards,gote2019git,ichimura2015analysis,ohira2005supporting,schwind2010tool,valetto2007using} \\
\hline
\end{tabular}
\caption{Number of people that are inside the \acp{DSN}. Missing means that the number of people is not or not accurately reported in the publication, NA means that the publication did not conduct an empirical study.}
\label{tbl:numdevs}
\end{table}

\begin{mdframed}
\textbf{Answer to RQ\,3b:} 
Most publications report networks that have more than 100 vertices. The number of developers is often much larger than the number of projects, because large-scale projects with big communities are analyzed. 
\end{mdframed}

\subsection{Influential Publications}
\label{sec:influential-publications}

We collected data regarding the citation counts from Google Scholar. We take the pattern from the ACM Distinguished Paper awards to define our criterion for influential publications, and consider the top 10\% with the most citations as influential. Since we have 255 publications, this means we consider the 25 publications with the most citations (Table~\ref{tbl:mostcited}). We note that the citations for the third most cited paper~\citeS{bird2006mining} also include the citations for the paper~\citeS{bird2006mining1}, because the two publications are considered as the same paper by Google Scholar. The 25 most influential publications address 
\begin{itemize}
    \item software development with globally distributed project members~\citeS{herbsleb2003empirical,ahuja2003individual,hinds2006structures};
    \item community structures in software development projects~\citeS{bird2006mining,crowston2005social,ducheneaut2005socialization,madey2002open,bird2008latent,crowston2006hierarchy,lopez2004applying,xu2005topological};
    \item the formation of teams in projects through collaboration~\citeS{hahn2008emergence,crowston2007self};
    \item the identification of relationships between developers~\citeS{begel2010codebook};
    \item the impact of coordination requirements between developers on tool design~\citeS{cataldo2006identification} and modularization~\citeS{cataldo2008socio};
    \item communication issues~\citeS{damian2007awareness} and trust~\citeS{sarker2011role}; 
    \item the identification of core developers~\citeS{crowston2006core};
    \item predictions to support software engineering processes, i.e., bug triage~\citeS{jeong2009improving}, defect prediction~\citeS{pinzger2008can,meneely2008predicting,bird2009putting}, build failure prediction \citeS{wolf2009predicting}, and collobariotions~\citeS{mcdonald2003recommending}. 
\end{itemize}

\begin{longtable}{p{4.7cm}p{3.8cm}p{1cm}>{\raggedleft\arraybackslash}p{1cm}}
\hline
\textbf{Title} & \textbf{Authors} & \textbf{Year} & \textbf{\#Cit.} \\
\hline\hline
An empirical study of speed and communication in globally distributed software development & James D. Herbsleb, Audris Mockus & 2003 & 1127 \\ \hdashline
Individual Centrality and Performance in Virtual R\&D Groups: An Empirical Study & Manju K. Ahuja, Dennis F. Galletta, Kathleen M. Carley & 2003 & 665 \\ \hdashline
Mining email social networks & Christian Bird, Alex Gourley, Premkumar Devanbu, Michael Gertz, Anand Swaminathan & 2006 & 644 \\ \hdashline
The social structure of free and open source software development & Kevin Crowston, James Howison & 2005 & 602 \\ \hdashline
Identification of Coordination Requirements: Implications for the Design of Collaboration and Awareness Tools & Marcelo Cataldo, Patrick A. Wagstrom, James D. Herbsleb, Kathleen M. Carley & 2006 & 465 \\ \hdashline
Socialization in an Open Source Software Community: A Socio-Technical Analysis & Nicolas Ducheneaut & 2005 & 459 \\ \hdashline
Improving Bug Triage with Bug Tossing Graphs & Gaeul Jeong, Sunghun Kim, Thomas Zimmermann & 2009 & 434 \\ \hdashline
The Open Source Software Development Phenomenon: An Analysis Based on Social Network Theory & Gregory Madey, Vincent Freeh, Renee Tynan & 2002 & 342 \\ \hdashline
The role of communication and trust in global virtual teams: A social network perspective & Saonee Sarker, Manju K. Ahuja, Suprateek Sarker, Sarah Kirkeby & 2011 & 313 \\ \hdashline
Latent social structure in open source projects & Christian Bird, David Pattison, Raissa D'Souza, Vladimir Filkov, Premkumar Devanbu & 2008 & 301 \\ \hdashline
Socio-Technical Congruence: A Framework for Assessing the Impact of Technical and Work Dependencies on Software Development Productivity & Marcelo Cataldo, James D. Herbsleb, Kathleen M. Carley & 2008 & 300 \\ \hdashline
Emergence of New Project Teams from Open Source Software Developer Networks: Impact of Prior Collaboration Ties & Jungpil Hahn, Jae Y. Moon, Chen Zhang & 2008 & 296 \\ \hdashline
Can developer-module networks predict failures? & Martin Pinzger, Nachiappan Nagappan, Brendan Murphy & 2008 & 243 \\ \hdashline
Predicting failures with developer networks and social network analysis & Andrew Meneely, Laurie Williams, Will Snipes, Jason Osborne & 2008 & 241 \\ \hdashline
Self-organization of teams for free/libre open source software development & Kevin Crowston, Qing Li, Kangning Wei, U. Y. Eseryel, James Howison & 2007 & 240 \\ \hdashline
Recommending collaboration with social networks: A comparative evaluation & David W. McDonald & 2003 & 226 \\ \hdashline
Codebook: discovering and exploiting relationships in software repositories & Andrew Begel, Yit P. Khoo, Thomas Zimmermann & 2010 & 222 \\ \hdashline
Predicting build failures using social network analysis on developer communication & Timo Wolf, Adrian Schröter, Daniela Damian, Thanh H.D. Nguyen & 2009 & 217 \\ \hdashline
Awareness in the Wild: Why Communication Breakdowns Occur & Daniela Damian, Luis Izquierdo, Janice Singer, Irwin Kwan & 2007 & 214 \\ \hdashline
Structures that work: social structure, work structure and coordination ease in geographically distributed teams & Pamela Hinds, Cathleen McGrath & 2006 & 209 \\ \hdashline
Applying social network analysis to the information in CVS repositories & Luis Lopez-Fernandez, Gregorio Robles, Jesus M. Gonzales-Barahona & 2004 & 207 \\ \hdashline
Core and Periphery in Free/Libre and Open Source Software Team Communications & Kevin Crowston, Kangning Wei, Qing Li, James Howison & 2006 & 204 \\ \hdashline
Hierarchy and centralization in free and open source software team communications & Kevin Crowston, James Howison & 2006 & 200 \\ \hdashline
A Topological Analysis of the Open Source Software Development Community & Jin Xu, Yongqin Gao, Scott Christley, Gregory Madey & 2005 & 198 \\ \hdashline
Putting It All Together: Using Socio-technical Networks to Predict Failures & Christian Bird, Nachiappan Nagappan, Harald Gall, Brendan Murphy, Premkumar Devanbu & 2009 & 197 \\ \hdashline
\hline
\hline
\caption{Top 10\% of publications ranked by number of citations. Data according to Google Scholar collected on 2020-02-24.}
\label{tbl:mostcited}
\end{longtable}

\begin{mdframed}
\textbf{Answer to RQ\,4a:} 
There are many publications on \acp{DSN} with a high citation count. The most influential publications address a very diverse number of topics, which highlights that there are many use cases for \acp{DSN} in software engineering research.
\end{mdframed}

\subsection{Influential Authors}

We identified 481 different authors who contributed to the literature on \acp{DSN}. We use a biblometric approach to identify the most influential of these authors, based on three different indicators: 1) the number of citations of all publications on \acp{DSN}; 2) the number of publications on \acp{DSN};  and 3) the number of publications on \acp{DSN} we identified as influential (Section~\ref{sec:influential-publications}). We consider the top-5 authors in each category to be the most influential. For the bibliometric data we collected, this means that an author has to have at least 1397 citations, 8 publications, or 3 influential publications to be considered as one of the most influential authors. 

Table~\ref{tbl:authors} shows the nine most influential authors we identified according to these criteria. Below, we briefly summarize the research directions of the influential authors. We discuss authors that frequently collaborated with each other as a group. 
\begin{itemize}
    \item James D. Herbsleb and Kathleen M. Carley are co-authors of three influential publications as well as several other publications. Herbsleb and Carley are both professors at Carnegie Mellon University. Their work covers structures and collaboration in global software engineering as well as socio-technical congruence within projects.
    \item Premkumar Devanbu was the PhD advisor of Christian Bird, who wrote his dissertation on \acp{DSN}. Their work addressed social structures and openness of open source projects, as well as build failure prediction.
    \item Kevin Crowston was the PhD advisor of James Howison, who wrote his dissertation on \acp{DSN}. Their work addressed community structures for open source software development.
    \item Daniela Damian collaborated with different authors as part of her work on communication between developers from different perspectives.
    \item Gregory Madey was the lead author of the first paper on \acp{DSN} we identified. He enabled many early papers through the SourceForge Research Data Archive~\cite{van2008advances}.
    \item Vladimir Filkov contributed to different aspects, including homophily, developer initiation into projects, communication behavior, as well as general structural aspects of \acp{DSN}.
\end{itemize}

\begin{table}
\begin{tabular}{p{3.5cm}>{\raggedleft\arraybackslash}p{1.7cm}>{\raggedleft\arraybackslash}p{1.7cm}>{\raggedleft\arraybackslash}p{3.5cm}}
\hline
\textbf{Author} & \textbf{\#Cit.} & \textbf{\#Pubs.} & \textbf{\#Influential Pubs.} \\
\hline\hline
James D. Herbsleb & 2271 & 7 & 3 \\
Kathleen M. Carley & 1488 & 4 & 3 \\
Premkumar Devanbu & 1477 & 10 & 3 \\
Christian Bird & 1472 & 8 & 3 \\
James Howison & 1397 & 6 & 4 \\
Kevin Crowston & 1397 & 6 & 4 \\
Daniela Damian & 1026 & 11 & 2 \\
Gregory Madey & 704 & 9 & 2 \\
Vladimir Filkov & 498 & 9 & 1 \\
\hline
\end{tabular}
\caption{Most influential authors according to the number of citations, number of publications, and number of influential publications.}
\label{tbl:authors}
\end{table}

\begin{mdframed}
\textbf{Answer to RQ\,4b:}
We identified 9 out of 481 authors as highly influential. The most influential author is James D. Herbsleb with over 2200 citations and 7 publications within the field of \acp{DSN}, three of which are among the top 10\% of all publications with respect to the number of citations. 
\end{mdframed}

\subsection{Important Venues}

The identified papers were published in 118 different venues, i.e., journals, conferences, and workshops. Table~\ref{tbl:venues} lists the venues at which most papers on \acp{DSN} were published. Three conferences stand out: the International Conference on Open Source Software (OSS), the International Conference on Software Engineering (ICSE), and the International Conference on Mining Software Repositories (MSR). 18\% of all papers on \acp{DSN} were published at these three venues. This is not surprising, as most publications analyze open source projects or ecosystems and employ software repository mining techniques. The ICSE is the top conference in the software engineering field, which highlights that there are papers of outstanding quality on \acp{DSN}. We note that the venues with most publications are mostly conferences, which is in line with the general conference-centered publication system of computer science research. The only three journals that made it into this list are Empirical Software Engineering, Information and Software Technology and the Journal of Systems and Software. However, there are also publications in other premier journals, e.g., the IEEE Transactions on Software Engineering~\citeS{kwan2011does,herbsleb2003empirical,lim2012stakerare,catolino2019gender}, ACM Transactions on Software Engineering Methodology~\citeS{singh2010small}, Management Information Systems Quarterly~\citeS{singh2011network}, and PLOS ONE~\citeS{xuan2016converging,Sapkota2020}.

\begin{table}
\begin{tabular}{p{9.9cm}>{\raggedleft\arraybackslash}p{1.3cm}}
\hline
\textbf{Venue} & \textbf{\#Pubs.}\\
\hline\hline
International Conference on Software Engineering (ICSE) & 17 \\
International Conference on Open Source Software (OSS) & 16 \\
International Conference on Mining Software Repositories (MSR) (Workshop until 2007, Working Conference until 2015) & 15\\
Conference on Computer Supported Cooperative Work (CSCW) & 10 \\
International Conference on the Foundations of Software Engineering (FSE) & 8 \\
Hawaii International Conference on System Sciences (HICSS) & 8 \\
Asia-Pacific Software Engineering Conference (APSEC) & 8 \\
Empirical Software Engineering, Springer & 7 \\
International Workshop on Cooperative and Human Aspects of Software Engineering (CHASE) & 7 \\
Information and Software Technology, Elsevier & 6 \\
Journal of Systems and Software (JSS) & 5 \\
International Conference on Global Software Engineering (ICGSE) & 5 \\
International Conference on Software Maintenance and Evolution (ICSME) (ICSM until 2013) & 5 \\
\hline
\end{tabular}
\caption{Most important publication venues determined by the number of papers published. We omitted labels like IEEE, ACM, or similar from the conference names, as they often changed slightly throughout the years.}
\label{tbl:venues}
\end{table}

\begin{mdframed}
\textbf{Answer to RQ\,4c:} 
The papers on \acp{DSN} were published in 118 different venues, including journals, conferences, and workshops. The most prominent venues are the ICSE, the OSS, and the MSR. Only three journals are in the list of the most important venues, which highlights that most \acp{DSN} research is published at conferences. 
\end{mdframed}

\subsection{Interest over Time}

\begin{figure}
    \centering
    \includegraphics[width=0.8\textwidth]{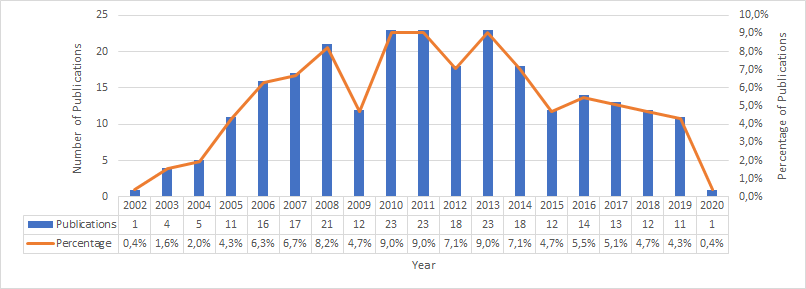}
    \caption{Publications per year.}
    \label{fig:pubs_per_year}
\end{figure}

Another interesting aspect is the interest of researchers with respect to \acp{DSN} over time measured by the number of publications per year. Figure~\ref{fig:pubs_per_year} depicts the total number of publications per year since the initial publication by Madey\etal~\citeS{madey2002open} in 2002. The topic quickly gained traction in the research community with rising numbers of publications until the interest became steady with 11 to 21 publications per year between 2005 and 2018. However, there seems to be a slight decline in the interest in \acp{DSN} since 2014. We note that due to the time of our search, the data for 2020 (and possibly 2019) is incomplete.

\begin{figure}
    \centering
    \includegraphics[width=0.98\textwidth]{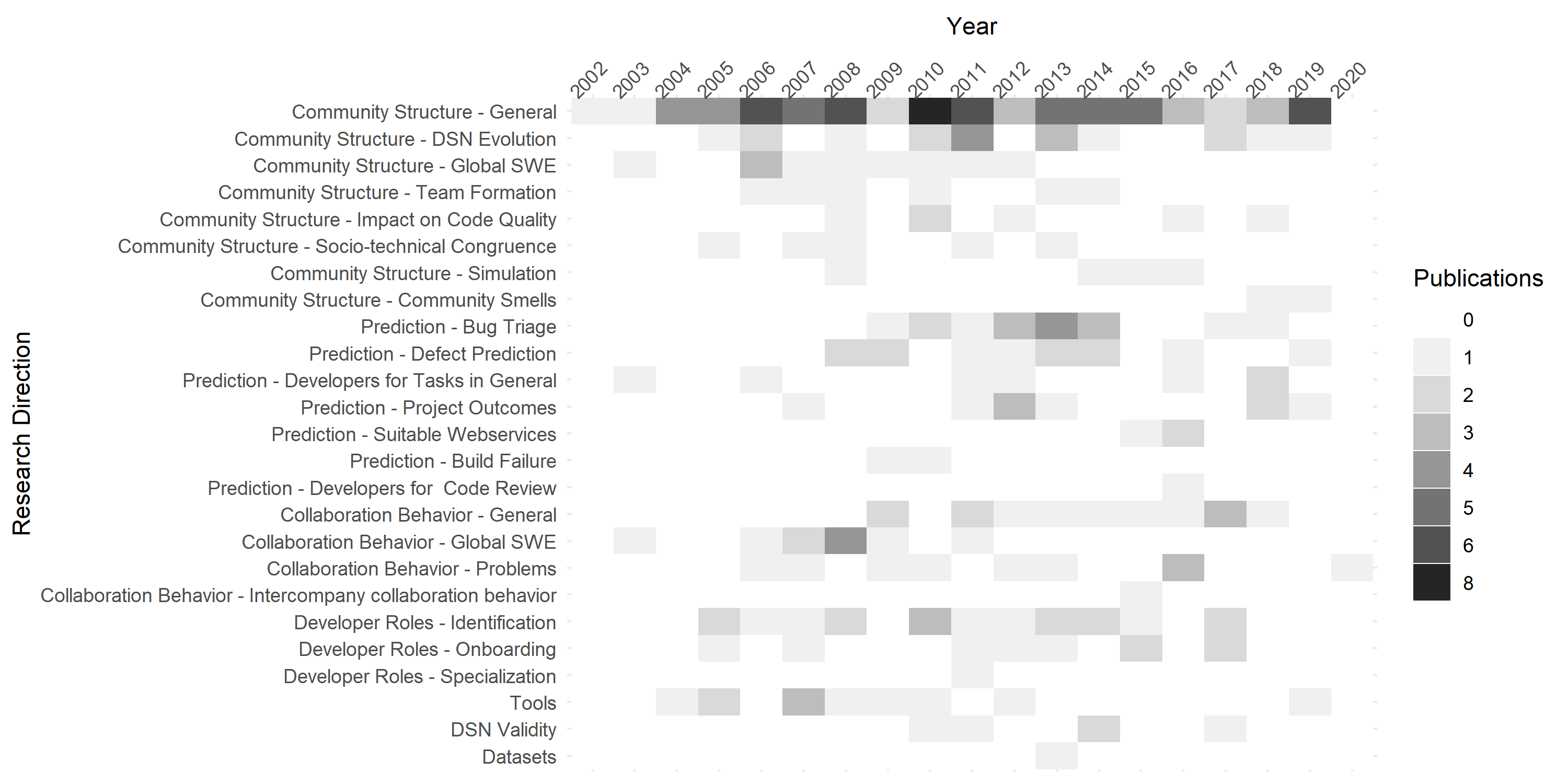}
    \caption{Research Directions over Time}
    \label{fig:research_over_time}
\end{figure}

Figure~\ref{fig:research_over_time} shows a heat map of the research directions over time. The interest in the general structure of \acp{DSN} has been steadily high over time. For all research directions, we see a fairly steady interest in the years 2005 to 2013. From 2013 onwards, the decline in overall interest in \acp{DSN} is also reflected in a more erratic interest in specific research directions. The most notable decline since 2013 is regarding global software engineering and tools, for which there are no publications anymore, even though the interest before was fairly high. Another important aspect is regarding the recent research, i.e., the years 2018 onwards. We note that the diversity of the research topics for which \acp{DSN} are used has declined, i.e., the ongoing research currently seems to be focused on community structures and predictive usages of \acp{DSN}. However, we see that there are still new topics emerging, i.e., the analysis of community smells~\citeS{palomba2018beyond,catolino2019gender}.

\begin{mdframed}
\textbf{Answer to RQ\,5:} 
The interest in \acp{DSN} peaked from 2006 till 2013. Regardless, there is still a steady interest in the topic, albeit we note a decrease in the diversity of research topics since 2018. 
\end{mdframed}
\section{Discussion of Open Issues}
\label{sec:discussion}

Our mapping study shows that \acp{DSN} are a versatile method for software engineering research. Mostly, they are used for the analysis of social structures and communication. However, the applications of \acp{DSN} range beyond that, e.g., for predictive purposes. Within this section, we discuss open problems in \ac{DSN} research. 

\subsection{General Issues}

Here, we discuss general issues within the current body of work on \acp{DSN}, that should be addressed by future work. 

\subsubsection{Lack of Guidelines}

There are no guidelines on how to conduct \ac{DSN} research. Therefore, the studies on \acp{DSN} are performed and described very heterogeneously. This is not an issue in itself, as heterogeneity can also be positive if different aspects are analyzed. Moreover, many publications perform well-designed case studies and report all important data regardless of the lack of guidelines. However, we observed several issues that result from the inconsistent way studies with \acp{DSN} are performed:
\begin{itemize}
    \item lack of reporting of the exact data sources and/or selection criteria for case study subjects;
    \item lack of reporting of important meta data about the study, e.g., number of projects, number of people; and
    \item lack of reporting of pre-processing steps performed with the data, e.g., to merge identities in case the same people used multiple aliases.
\end{itemize}

The development of guidelines for research on \acp{DSN} can, therefore, help to enhance the quality of \ac{DSN} research in general. 

\subsubsection{Studies with High External Validity}

\begin{figure}
    \centering
    \includegraphics[width=0.7\textwidth]{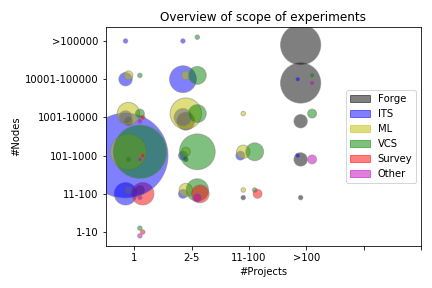}
    \caption{Relationship between study sizes and data sources}
    \label{fig:size-vs-datasource}
\end{figure}

Our data shows that many results regarding \acp{DSN} were obtained only on very few projects, i.e., over 69\% of the publications used less than 11 projects to conduct their research. While this does not mean that the results are wrong or would not generalize to other contexts, this poses a threat to the generalizability of results, as such small samples can only represent populations with a limited context. This problem is to some degree further aggravated, because there is an overlap in the data that is used, i.e., multiple studies using the same data, sometimes the same single project (e.g., IBM Jazz or the Global Studio Project). Moreover, we noted a strong relation between the data sources and the size of studies. Figure~\ref{fig:size-vs-datasource} shows the size of the studies with relation to the data source. The larger circles mean more publications. Almost all publications with large numbers of people and projects were based on data from forges. Thus, an open issue considered for all future publications is to use larger sample sizes regarding the number of projects, to enable a better generalizability of results. This could either be done by harnessing data from forges or by collecting data for more projects from other data sources. 

\subsubsection{Lack of Replications}

There is general lack of replications in \ac{DSN} research. The publications are more or less independent of each other, the exception being multiple publications by the same authors building on each other. We did not find any study that explicitly tried to replicate prior results. The lack of replications is especially problematic due to the often very small numbers of projects considered (see above). Thus, we believe that replication studies on \ac{DSN} research are required for all research directions so far. 

\subsection{Open Topics}

Here, we discuss potential future directions of \ac{DSN} research. 

\subsubsection{Inter-company Collaborations}

Since more and more companies contribute to open source software and/or develop their own software products as open source, the collaboration between developers of competing companies becomes an important issue. If developers from competing organizations contribute to the same project, this could lead to issues within a project, that could be analyzed through \acp{DSN}, e.g., with respect to team formation, onboarding, collaboration problems, and even impacts on the socio-technical congruence of projects. Within our mapping study, we only discovered one publication in this direction~\citeS{teixeira2015lessons}.

\subsubsection{DSNs from Multiple Sources}

The use of multiple sources for \ac{DSN} studies allows a deeper analysis of developer communities. For example, how does the community on a \ac{ML} differ from the community that can be observed in pull request discussions or in an \ac{ITS}? Can we infer something about onboarding of developers from their integration in different \acp{DSN}? Do projects that use an \ac{ITS} and a \ac{ML} exhibit different collaboration properties than projects that just use an \ac{ITS} or a \ac{ML}? What exactly is the temporal-spatial relationship between the \ac{DSN} structures of different sources? Does research regarding the team formation of projects based on \acp{ML} yield the same results as research on team formation on \acp{ITS}? How does migration to a new \ac{ITS} affect the community structure? All of these are currently open questions. Especially the comparison of \acp{DSN} that are based on different data sources has been neglected so far, with only a single publication that directly compares the \ac{DSN} structure obtained from \ac{ITS} data with that obtained from \ac{VCS} data~\citeS{aljemabi2017empirical}. 

\subsubsection{Diversity in DSNs}

The role of gender and other issues related to diversity is an important recent trend in current software engineering research. There is already one recent publication that touches on the relation between gender and DSNs~\citeS{catolino2019gender}. We believe that insights into the question if and how gender or other diversity related features impact DSNs can give researchers and practioneers valuable insights that may help to make software engineering both more inclusive and more effective. 

\subsubsection{Applications using DSNs}

The current literature on \acp{DSN} has a strong focus on understanding community structures and the implications of the community structure on issues like developer roles, team formation, and collaboration behavior. However, there are only relatively few actionable applications of \acp{DSN}. CodeBook~\citeS{begel2010codebook} is a notable exception that demonstrates how \acp{DSN} can be used to improve the daily life of software developers. While other publications also study applications of \acp{DSN}, e.g., for defect prediction, failure prediction, or developer recommendations, they are mostly not accompanied by a tool that makes the research actionable for practitioners. The tool papers that we identified cover mostly the visualization of \acp{DSN}. While visualizations are a useful tool for the analysis of communities, they are not actionable applications of \acp{DSN}. We believe that research that produces actionable tools can have a big impact, e.g., on already considered issues like bug triage or developer recommendations. 

\subsubsection{Data sets}

We only identified a single publication that published a \ac{DSN} as data set. While there are other publications that are based on public data sets, e.g., the source forge dump~\cite{van2008advances}, these data sets are not yet \acp{DSN}. They only contain the data necessary to create a \ac{DSN}. While there are certainly use cases, in which new \acp{DSN} must be created, e.g., because different information is used to create links between developers, there are also cases for which dedicated data sets on \acp{DSN} would have advantages. For example, benchmark data sets could allow, e.g., to compare different approaches for developer recommendation or the identification of core developers. Moreover, the collection of data from a large amount of software repositories can be very time consuming. Data sets for a large amount of projects could help with this issue, and, e.g., enable larger studies with \acp{ML} as sources for projects. 

\section{Related Work}
\label{sec:relatedwork}

Our systematic mapping study is not the first literature study that covers \acp{DSN}. Within this section, we discuss related literature studies on \acp{DSN}, their differences to our work, and how we utilized them as sanity checks for our work. 

Closest to our work is the survey by Zhang\etal~\cite{zhang2014developer}. Similar to our work, the authors analyzed the data sources, as well as topics that were addressed with \acp{DSN}. However, there are several notable differences between the work by Zhang\etal{} and our work. First, the search strategy by Zhang\etal{} is different from ours. They used the search term "developer network" and identified 20 publications related to \acp{DSN} within the first 50 hits on Google Scholar. Using these publications as seed, the authors performed one round of forward/backward snowballing and identified a total of 86 primary studies this way. Due to this limited search, the authors only identified a limited body of the relevant literature. In comparison, we use more search terms and multiple search engines, consider 750 instead of 50 hits per search term/seach engine, and performed exhaustive backward and forward snowballing until no further papers were identified, and performed an exhaustive search with Scopus. Moreover, the focus of the presentation and focus of the work from Zhang\etal{} differs from ours. We provide a systematic mapping of approaches to topics through inductive coding, i.e., are interested in the general topics and trends that are analyzed. In comparison, Zhang\etal{} provide a more detailed description of different approaches to address research topics, but do not systematically map publications into different categories. However, we used the description of research topics described by Zhang\etal{} as starting point for our inductive coding, but identified additional topics, e.g., global software engineering, team formation, inter-company collaboration behavior, and developer onboarding. Another difference to our work is that Zhang\etal{} also report on the metrics that were used for the analysis of the \ac{DSN}, an aspect that is not covered by our mapping study. 

The literature study by Tamburri\etal~\cite{tamburri2013organizational} uses grounded theory to identify different types of social structures within open source software development. Thus, their focus is different from ours, which is on \acp{DSN} in general, not on social structures. However, \acp{DSN} play an important role in the study by Tamburri\etal{} and are part of the literature that they identify. Our inductive coding approach is in principle similar to the approach used by Tamburri\etal, in the sense that we inductively infer the relevant concepts from the identified publications. However, the goal was completely different, i.e., we wanted to identify research topics instead of types of organizational structures. Thus, while we identify broad topics of research, Tamburri\etal{} identified detailed information for the topic of social and collaboration structures. Due to the different focus, the search strategies also differ. Most importantly, the search by Tamburri\etal{} also covers search terms like ''organizational``, ''knowledge community`` and similar to account for the different focus. Moreover, the search engines used are different from ours. They used Web of Science, EBSCO, JSTOR knowledge storage, Wiley InterScience and ProQuest in addition to the search engines we used. On the other hand, we used Google Scholar, which was not used by Tamburri\etal. The authors identified 143 publications for their study. 

Manteli\etal~\cite{manteli2012adopting} performed a literature study to analyze \acp{DSN} with respect to global software development. Their focus was on coordination, cooperation, and communication aspects of global software development. This scope of this survey is narrower than our mapping study of \acp{DSN} without further restrictions. Thus, while we identify broad topics of research, Manteli\etal{} identified detailed information for the \acp{DSN} in global software engineering. 
This shows in the difference in search terms and inclusion criteria. Moreover, there is a difference in search engines used. Manteli\etal{} used EBSCO and Wiley InterScience in addition to the engines we used, but did not use Google Scholar. The authors identified 23 primary studies on \acp{DSN} with a relation to global software development. 

Abufouda and Abukwaik~\cite{Abufouda2017} performed a systematic literature review on \acp{DSN} with the goal to identify how reliable constructed social networks are. This goal is different from our general focus, which shows, e.g., in the different exclusion criteria. The authors used the same search engines we also used, with the exception of Google Scholar and SCOPUS which were not considered. The authors identify 23 primary studies that meet the criteria for their survey. The data the authors collected is very detailed with respect to the required description of the model and covers aspects like vertex types, edge types, and validation criteria. Thus, the work by Abufouda and Abuwaik focuses on evaluating aspects related to the internal validity of studies. In comparison, we collect data related to the external validity of \ac{DSN} studies in our work, i.e., the scope of the analysis that is conducted. 

In addition to our comparison with related work above, there are several differences between our work and all the related literature. No other work performed a bibliometric assessment of influential authors, papers, and venues or the trends over time. Moreover, no work in the literature provides information about the scope of the networks, i.e., the number of projects and participants that are analyzed through \acp{DSN} in a publication. Therefore, none of the prior literature studies provides answers to our research questions RQ3-RQ5. Additionally, as discussed above, none of the works provides a complete or systematic mapping of research topics which could be used to answer RQ1 or empirical data to support the answer for RQ2. Overall, our work goes well beyond the currently available literature studies on \acp{DSN}, both in terms of identified publications as well as due to the research questions we address.

\section{Conclusion}
\label{sec:conclusion}

This article presents the results of our systematic mapping study on DSNs. We identified 255 primary studies published between 2002 and 2020. Our results show that DSNs were used for the analysis of many different software engineering research topics since their initial use in the year 2002~\citeS{madey2002open}, e.g., the analysis of community structures, the creation or improvement of prediction models, the study of collaboration behavior, and the identification of developer roles within projects. Moreover, we found that the social networks are often modelled based on data collected from a single repository, e.g., a forge like GitHub or SourceForge, version control systems like Git or SVN, issue tracking systems like Jira or Bugzilla, or mailing lists. A few notable exceptions use on-site observation centric techniques instead of data from software repositories. We observed a tendency that many publications only use a small set of sample projects, which may inhibit the generalizability of findings. Related to this is a general lack of replications within the body of work, i.e., we did not find a single replication study. Through a bibliometric assessment we found that there are many highly cited papers on DSNs on a diverse set of topics, which highlights the many use cases for DSNs in software engineering research. Our data shows that the interest in DSNs in research is still high, though there is a slight declining trend in recent years after the interest peaked in 2014. 

Based on our findings, we suggest that future research addresses aspects that were neglected so far, e.g., inter-company collaborations~\citeS{teixeira2015lessons}, practical applications of DSNs~\citeS{begel2010codebook}, and the relationship between social network structures and diversity~\citeS{catolino2019gender}. Moreover, we believe that replication studies can help to address the question if current results from the state of the art generalize beyond the often relatively small set of projects that were used in many publications.

\newpage
\bibliographystyle{elsarticle-num}
\bibliography{literature}

\bibliographystyleS{elsarticle-num}
\bibliographyS{primary_studies}

\end{document}